\documentclass[prd,twocolumn,a4paper,superscriptaddress]{revtex4}
\usepackage{graphicx,amssymb}
\begin{document}
\newcommand{\be}{\begin{equation}}
\newcommand{\ee}{\end{equation}}
\newcommand{\bq}{\begin{eqnarray}}
\newcommand{\eq}{\end{eqnarray}}
\newcommand{\bsq}{\begin{subequations}}
\newcommand{\esq}{\end{subequations}}
\newcommand{\bc}{\begin{center}}
\newcommand{\ec}{\end{center}}
\newcommand {\R}{{\mathcal R}}
\newcommand{\al}{\alpha}
\newcommand\lsim{\mathrel{\rlap{\lower4pt\hbox{\hskip1pt$\sim$}}
    \raise1pt\hbox{$<$}}}
\newcommand\gsim{\mathrel{\rlap{\lower4pt\hbox{\hskip1pt$\sim$}}
    \raise1pt\hbox{$>$}}}

\title{The Cosmological Evolution of Domain Wall Networks}
\author{J.C.R.E. Oliveira}
\email{jeolivei@fc.up.pt}
\affiliation{Centro de F\'{\i}sica do Porto, Rua do Campo Alegre 687, 4169-007 Porto, 
Portugal}
\affiliation{Departamento de F\'{\i}sica da Faculdade de Ci\^encias
da Universidade do Porto, Rua do Campo Alegre 687, 4169-007 Porto, 
Portugal}
\author{C.J.A.P. Martins}
\email[Electronic address: ]{C.J.A.P.Martins@damtp.cam.ac.uk}
\affiliation{Centro de F\'{\i}sica do Porto, Rua do Campo Alegre 687, 4169-007 Porto, 
Portugal}
\affiliation{Department of Applied Mathematics and Theoretical 
Physics,
Centre for Mathematical Sciences,\\ University of Cambridge,
Wilberforce Road, Cambridge CB3 0WA, United Kingdom}
\author{P. P. Avelino}
\email[Electronic address: ]{ppavelin@fc.up.pt}
\affiliation{Centro de F\'{\i}sica do Porto, Rua do Campo Alegre 687, 4169-007 Porto, 
Portugal}
\affiliation{Departamento de F\'{\i}sica da Faculdade de Ci\^encias
da Universidade do Porto, Rua do Campo Alegre 687, 4169-007 Porto, 
Portugal}

\date{26 October 2004}

\begin{abstract}
We have studied the cosmological evolution of domain wall networks in two, three and four spatial dimensions using high-resolution field theory simulations. The dynamical range and number of our simulations is larger than in previous works, but does not allow us to exclude previous hints of deviations to the naively expected scale-invariant evolution. These results therefore suggest that the approach of domain wall networks to linear scaling is a much slower process than that of cosmic strings, which has been previously characterized in detail. 
\end{abstract}
\pacs{98.80.Cq, 11.27.+d, 98.80.Es}
\keywords{Cosmology; Topological Defects; Domain Walls; Numerical Simulation}
\preprint{DAMTP-2003-44}
\maketitle

\section{\label{intr}Introduction}

The cosmological consequences of primordial phase transitions that are thought to have happened in the early universe have been the subject of many studies in recent years. One of these inevitable consequences is the formation of topological defects \cite{Kibble,Book}. While most such studies have focused on cosmic strings, domain walls can be of interest too, although observational constraints rule them out if their symmetry breaking scale is $\eta\ge 1MeV$ \cite{Zeldovich}. However, note that this constraint is on the assumption that they are topologically stable and that their cosmological evolution is scale-invariant (\textit{id est}, that the network correlation length is proportional to the horizon size, or equivalently that the total area density is
inversely proportional to it).

In this work we want to highlight the fact that, in spite of some simple analytic arguments in favor of it \cite{Hindmarsh}, this hypothetical scale-invariant solution has not yet received any strong backing from numerical simulations. In fact, in previous simulations of cosmological domain wall evolution \cite{Press,Coulson,Larsson,Fossils,Garagounis} some hints for deviations to a scale-invariant evolution were found. The question we wish to address presently is whether these deviations remain if one performs larger runs.

We report on a large set of high-resolution simulations of domain walls in two, three and four spatial dimensions, using the standard Press-Ryden-Spergel (PRS) algorithm \cite{Press}. Our boxes are typically larger (hence can be evolved for a longer dynamic range) that those in previous works. Gains in statistical accuracy also come from performing large numbers of runs. We highlight the differences between running a large number of small runs, or a small number of larger ones. We do find some statistical evidence for deviations to the scale-invariant behaviour, though there is some dependency on dimensionality and box sizes. While 2D and 3D numerical simulations have been previously carried out (though usually in smaller boxes or with smaller dynamic range) by a number of authors \cite{Press,Coulson,Larsson,Fossils,Garagounis}, we have also carried out for the first time a series of 4D simulations, which in a simple phenomenological way may be of interest to brane world scenarios \cite{tye0,Brax}.

In what follows we start with a brief description of the PRS algorithm and then proceed to present some tests of our code and the main results of our simulations. We also compare these with the work of previous authors, and highlight issues having to do with the slightly different methods used in each case. Finally, we briefly comment on the cosmological implications of our results.

\section{\label{prs}Evolution of the Domain Walls}

Let us consider the evolution of a domain wall network in a flat
homogeneous and isotropic Friedmann-Robertson-Walker (FRW) universe 
with line element
\begin{equation}
ds^{2}=dt^{2}-a^{2}(t)(dx^{2}+dy^{2}+dz^{2})\,,\label{metric}
\end{equation}
where $a(t)$, is the cosmological expansion factor, $x$, $y$ and
$z$ are co-moving coordinates and $t$ is physical time. The
dynamics of a scalar field $\phi$ is determined by the Lagrangian
density
\begin{equation}
\mathcal{L}={\frac{1}{2}}\phi_{,\alpha}\phi^{,\alpha}-V(\phi)\,,
\label{action1}
\end{equation}
where we will take $V(\phi)$ to be the generic $\phi^{4}$ potential
with two degenerate minima given by 
\begin{equation}
V(\phi)=V_{0}\left({\frac{\phi^{2}}{\phi_{0}^{2}}}-1\right)^{2}\,.
\label{potential}
\end{equation}
This obviously admits domain wall solutions. By varying the action
\begin{equation}
S=\int dt\int d^{3}x{\sqrt{-g}}\mathcal{L}\,,\label{action2}
\end{equation}
with respect to $\phi$ we obtain the field equation of motion: 
\begin{equation}
{\frac{{\partial^{2}\phi}}{\partial 
t^{2}}}+3H{\frac{{\partial\phi}}{\partial 
t}}-\nabla^{2}\phi=-{\frac{{\partial 
V}}{\partial\phi}}\,.\label{dynamics}
\end{equation}
where $\nabla$ is the Laplacian in physical coordinates and 
$H=(da/dt)/a$ is the Hubble parameter.

Following \cite{Press} we have modified the equations of motion in such a way that the co-moving thickness of the domain walls is fixed in co-moving coordinates. This has a small impact on the large scale dynamics of the domain walls and should not affect the quantities we want to measure for the purpose of testing scaling properties \cite{Press}, provided a minimum acceptable thickness is used---this will be discussed in Sect. \ref{tests}. On the other hand, it allows us to resolve the domain walls through the network evolution. For a discussion of analogous issues in the context of cosmic strings see \cite{Moore}.

Then equation (\ref{dynamics}) becomes: 
\begin{equation}
{\frac{{\partial^{2}\phi}}{\partial\eta^{2}}}+\alpha\left(\frac{d\ln 
a}{d\ln\eta}\right){\frac{{\partial\phi}}{\partial\eta}}-{\nabla}^{2}\phi=
-a^{\beta}{\frac{{\partial 
V}}{\partial\phi}}\,.\label{dynamics2}
\end{equation}
where $\eta$ is the conformal time and $\beta_{1}$ and $\beta_{2}$
are constants: $\beta=0$ is chosen in order to have constant
co-moving thickness and $\alpha=3$ to require that the momentum
conservation law of the wall evolution in an expanding universe is
maintained \cite{Press} In fact we have numerically tested this with
$\alpha=2$ and $\alpha=4$ and no significant differences were 
found in the quantities we want to measure in this context.

Equation (\ref{dynamics2}) was integrated using a standard 
finite-difference
scheme, with the value of the Hubble damping term
determined by numerical integration of the Friedmann equation:
\begin{equation}
\frac{\partial\eta}{\partial 
a}=\frac{1}{H_{0}}\frac{1}{a\sqrt{\Omega_{m0}a^{-1}+\Omega_{\Lambda0}a^{2}+
(1-\Omega_{m0}-\Omega_{\Lambda0})}}\,.\label{friedmanavancado2}
\end{equation}
The conformal time evolution of the co-moving correlation length of the 
network  
$\xi_{c} \equiv \frac{V}{A}$ 
was determined. Following \cite{Press} the co-moving area $A$
was determined by finding the neighbouring grid points where $\phi$
has different signs and adding in these cases an increment $\Delta A$
divided by a weighting factor corresponding to the spatial gradient
of $\phi$:
\begin{equation}
A=\Delta 
A\sum\omega\frac{\mid\nabla\phi\mid}{\mid\partial\phi/\partial 
x\mid+\mid\partial\phi/\partial y\mid+\mid\partial\phi/\partial 
z\mid}\,,
\label{area}
\end{equation}
where $\omega=1$ if the link crosses the wall and $\omega=0$ otherwise.
Finally, the ratio between the kinetic and potential energy energy of the 
$\phi$ field is roughly proportional to 
\begin{equation}
F\equiv\frac{1}{A}\sum_{i,j,k}\left(\frac{\partial\phi_{ijk}}
{\partial\eta}\right)^{2}\,.
\label{energiacinetica}
\end{equation}
This quantity is directly related to the root-mean squared 
velocity of the domain walls 
which is conserved if a scaling solution exists.
We assume the initial value of $\phi$ to be a random variable
between $-\phi_{0}$ and $+\phi_{0}$ and the initial value of 
$\partial\phi/\partial\eta$
to be zero. 
See \cite{Fossils} for further discussion of these issues.

\section{\label{tests}Simple Tests of the Code}

A total of several thousand simulations in two, three and four spatial dimensions were run for various box sizes and other initial conditions, as further detailed below. A number of observables were used as diagnostics for scaling. We start by describing the one that is closest to previous work (alternatives will be described subsequently). In each run we looked for the best fit to the power law
\begin{equation}
\rho_{w}=\eta^{\lambda-1}\,,
\label{fit1}
\end{equation}
where $\rho_{w}$ is the wall energy density and $\eta$ the conformal time, and thereby measured the exponent $\lambda$. In terms of the co-moving and physical correlation lengths of the network this corresponds to the following scaling laws, respectively
\begin{equation}
\xi_{c}=\eta^{1-\lambda}\,,
\label{fit2}
\end{equation}
\begin{equation}
\xi=t^{1-\lambda/3}\,.
\label{fit3}
\end{equation}
Note that since we will be looking for deviations to the scale-invariant behaviour, which corresponds to $\lambda=0$, these will be larger if we work in co-moving coordinates, since any deviation exponent will be larger (though measurement errors could also differ).

Each fit was made considering only the reliable dynamical range of the corresponding simulation. Note that in the beginning of the simulation the initial conditions will influence the evolution, and additionally the co-moving correlation length must be significantly larger than the wall thickness for the evolution to be sufficiently well defined. On the other hand, towards the end of the simulation the boundary conditions may become relevant. Because in a (nearly) scale-invariant regime the average wall velocity is expected to be approximately constant, our choice of the reliable period was based on the constancy of the kinetic to potential energy ratio, as in \cite{Press}. 

\begin{figure}
\includegraphics[width=3.5in]{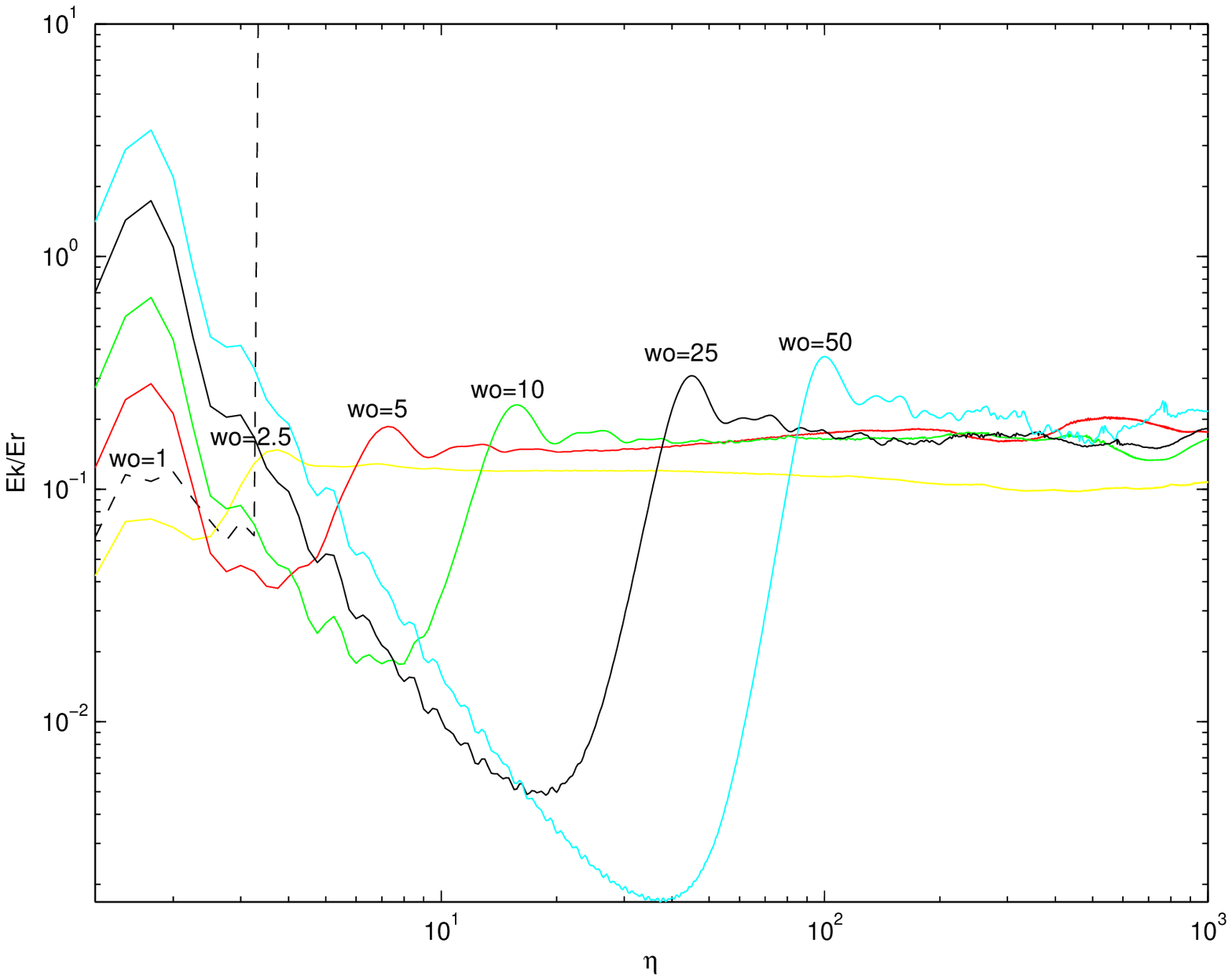}
\includegraphics[width=3.5in]{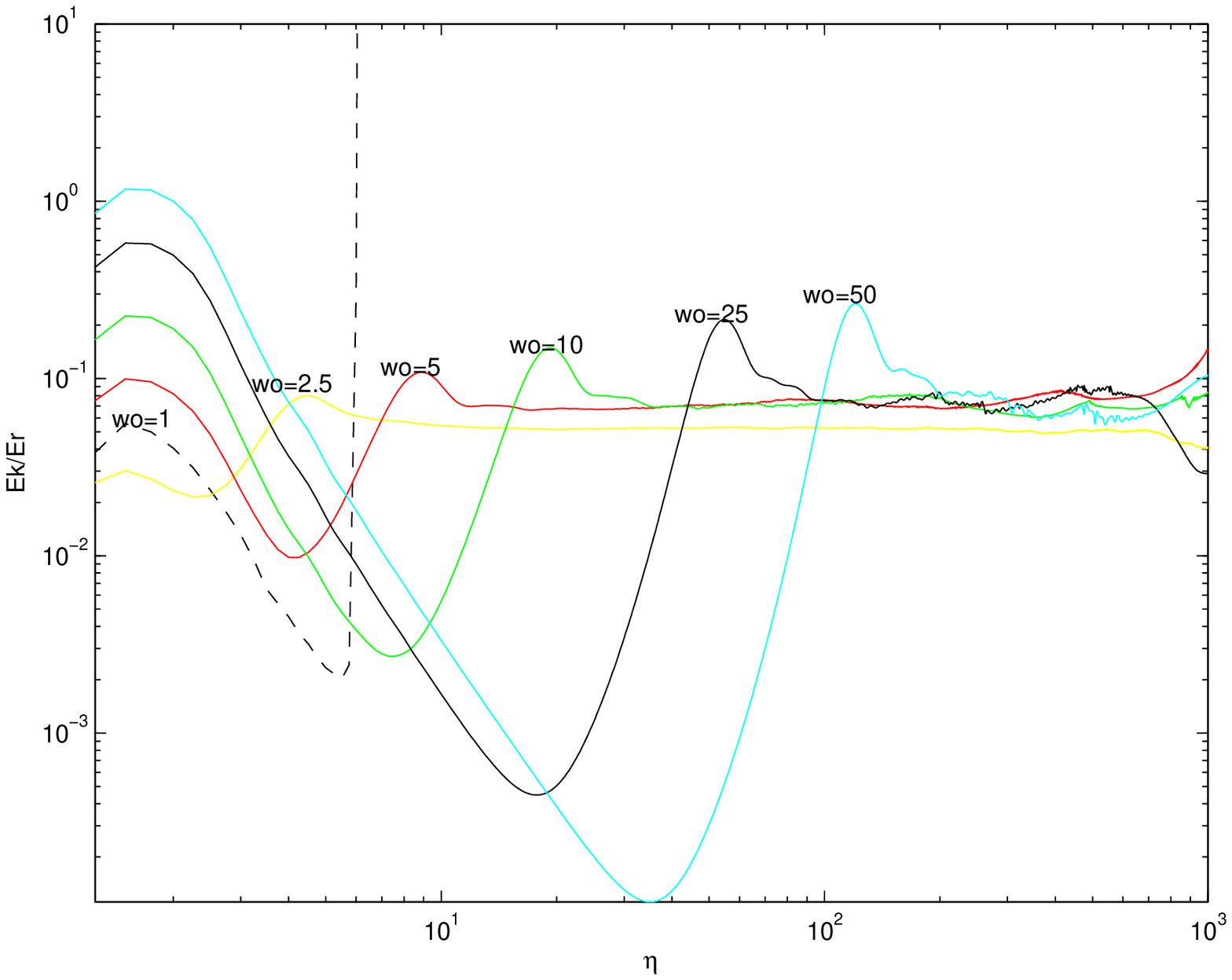}
\caption{\label{fig1}The evolution of the ratio between the kinetic and
potential energy energy of the $\phi$ field with conformal time for
$2048^{2}$ simulations in the radiation (top) and matter (bottom) epochs.
The simulations are run for a dynamical range of $1024$. $W_{0}$ is the wall thickness---notice that this sets the relaxation time scale.}
\end{figure}

One finds that early in the simulations this ratio displays strong 
oscillations while the initial conditions are relaxed. The relaxation timescale is typically given by the.wall thickness, which we parametrise by $W_0$ \cite{Press}. The ratio then stabilizes and there follows a period where it is approximately constant. Finally, towards the end of the simulations it may start to decrease or increase slightly. Two examples are shown in Fig. \ref{fig1}. We do our fits in this intermediate, near-constant regime, selecting the appropriate interval as follows. Firstly we aggressively choose a preliminary interval of conformal times, and determine the corresponding average value of the ratio. Secondly, if we find that at any point within the interval the kinetic energy varies by more than about ten percent, we reduce the interval (at either end, or both ends if necessary) until this is no longer so, and this final range is where the fit is then done.

\begin{figure}
\includegraphics[width=3.5in]{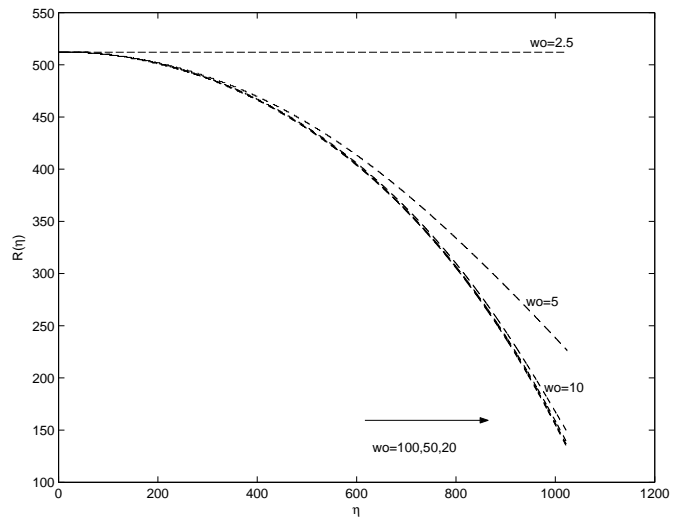}
\caption{\label{fig2}The collapse of spherical domain walls with different thickness $W_0$. All simulations have a dynamical range of $1200$.}
\end{figure}

Let us pause at this point to comment on the influence of the wall thickness on the dynamics of the networks. Consider first the collapse of a single spherical
domain wall of a given size, for different values of the thickness $W_0$. As we can see in Fig. \ref{fig2}, there is a minimum acceptable resolution of about 5 grid points (which will still yield inaccurate timescales), with 10 grid points being adequate. This can be confirmed by looking at network evolution, \textit{e.g.} of $2D$ boxes in the radiation and matter epochs as in Fig. \ref{fig1}. Here the relevant figure of merit is the average ratio between the kinetic and potential energies after the initial period of relaxation.

\begin{figure}
\includegraphics[width=3.5in]{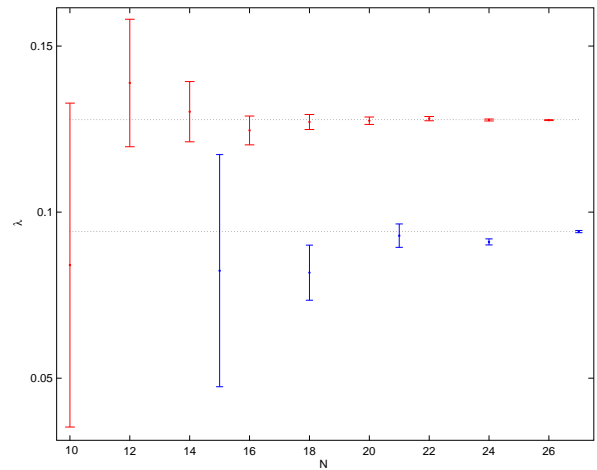}
\caption{\label{fig3}Measuring scaling exponents $\lambda$ for relatively small 2D (top set of points) and 3D (bottom set of points) simulations of domain wall networks. In each case $2^{N}$ is the total number of points in a box, and the error bar is the standard deviation in an ensemble of 100 simulations.}
\end{figure}

\begin{figure}
\includegraphics[width=3.5in]{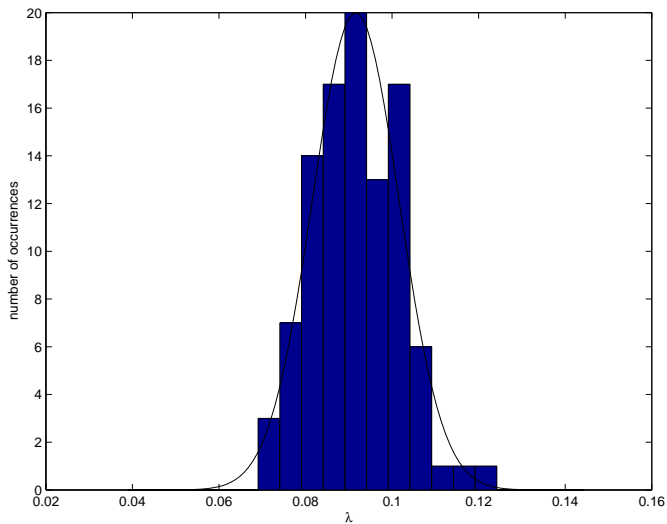}
\caption{\label{fig4}The error function distribution for an ensemble of 100 $256^{3}$ matter era runs. The thin line is the Gaussian function distribution.}
\end{figure}

We have carried out series of 100 test runs with different 2D and 3D boxes, to study the effect of the box size on the evolution of the networks. In each case we ran a number of different simulations, calculated an exponent for each simulation, and then computed the standard deviation for $\lambda$ from the ensemble of boxes (the fit range used was the same for all simulations with the same dynamical range). As shown in Fig. \ref{fig3}, the various results for the scaling exponent are consistent with each other, with the standard deviation in the result decreasing as we increase the box size. Also, as can be seen from the example in Fig. \ref{fig4}, the error function distribution for each ensemble of runs is approximately Gaussian (results for other series of runs are qualitatively similar). Notice that the exponents differ in 2D and 3D---we shall return to this point in the main results section.

Another simple test is to consider the evolution of the standard deviation in conformal time. In the scaling regime we would expect the standard deviation to be proportional to $\sqrt{N}/N\varpropto\eta^{3/2}$, $N=\left(L_{c}/\eta\right)^{3}$ and $L_{c}$ is the co-moving box size and $\eta$ is the horizon size. If one allows for the the correction $\eta^{-3/2\left(1-\lambda\right)}$, with $\lambda$ calculated as before for all the simulation fit range, we found that the standard deviation was almost constant and independent of conformal time.

\section{\label{numsim}Numerical Simulation Results}

\begin{table*}
\begin{tabular}{|c|c|c|c|c|c|c|c|c|c|c|c|}
\hline 
Dimension&
Reference&
Box Size&
Epoch&
$\eta_0$&
$\frac{\Delta \eta}{\eta_0}$&
$\frac{\Delta x}{\eta_0}$&
$W_0$&
Full Range&
Fit Range&
Runs&
Scaling exponent $\lambda$\tabularnewline
\hline 
\hline 
2D&
\cite{Press}&
$1024^2$&
Mat.&
$1.0$&
$1.0$&
$1.0$&
10&
512&
$5.0$&
10&
$0.129\pm0.015$\tabularnewline
\hline
2D&
\cite{Coulson}&
$1024^2$&
Rad.&
$1.0$&
$1.0$&
$1.0$&
5&
512&
$10.0$&
5&
$0.12\pm0.04$\tabularnewline
\hline
2D&
\cite{Larsson}&
$1024^2$&
Mat.&
$1.0$&
$1.0$&
$1.0$&
?&
512&
$10.0$&
few&
$0.05\pm0.08$\tabularnewline
\hline
2D&
\cite{Garagounis}&
$1024^2$&
Rad.&
$1.0$&
$0.1$&
$0.3$&
3.3&
1536&
$5.0$&
5&
$0.004\pm0.018$\tabularnewline
\hline
2D&
\cite{Garagounis}&
$1024^2$&
Mat.&
$1.0$&
$0.1$&
$0.3$&
3.3&
1536&
$5.0$&
5&
$0.008\pm0.014$\tabularnewline
\hline\hline
2D&
This work&
$4096^2$&
Rad.&
$1.0$&
$0.25$&
$1.0$&
10&
2048&
200&
100&
$0.0448\pm0.0045$\tabularnewline
\hline
2D&
This work&
$4096^2$&
Mat.&
$1.0$&
$0.25$&
$1.0$&
10&
2048&
200&
100&
$0.0340\pm0.0046$\tabularnewline
\hline
2D&
This work&
$8192^2$&
Rad.&
$1.0$&
$0.25$&
$1.0$&
10&
4096&
500&
10&
$0.0285\pm0.0067$\tabularnewline
\hline
2D&
This work&
$8192^2$&
Mat.&
$1.0$&
$0.25$&
$1.0$&
10&
4096&
500&
10&
$0.0095\pm0.0035$\tabularnewline
\hline\hline
3D&
\cite{Press}&
$200^3$&
Mat.&
$1.0$&
$1.0$&
$1.0$&
10&
100&
$2.0$&
9&
$0.08\pm0.02$\tabularnewline
\hline
3D&
\cite{Coulson}&
$128^3$&
Rad.&
$1.0$&
$1.0$&
$1.0$&
5&
64&
$6.4$&
5&
$0.11\pm0.06$\tabularnewline
\hline
3D&
\cite{Larsson}&
$128^3$&
Mat.&
$1.0$&
$1.0$&
$1.0$&
?&
64&
$6.4$&
few&
$0.08\pm0.08$\tabularnewline
\hline
3D&
\cite{Garagounis}&
$512^3$&
Rad.&
$1.0$&
$0.1$&
$0.3$&
3.3&
768&
$5.0$&
5&
$0.006\pm0.013$\tabularnewline
\hline
3D&
\cite{Garagounis}&
$512^3$&
Mat.&
$1.0$&
$0.1$&
$0.3$&
3.3&
768&
$5.0$&
5&
$0.003\pm0.012$\tabularnewline
\hline
3D&
\cite{Garagounis}&
$4096^3$&
?&
$1.0$&
$0.08$&
$0.3$&
3.3&
7680&
$5.0$&
1&
$0.015\pm0.003$\tabularnewline
\hline\hline
3D&
This work&
$256^3$&
Rad.&
$1.0$&
$0.25$&
$1.0$&
10&
128&
$12.0$&
100&
$0.0811\pm 0.0044$\tabularnewline
\hline
3D&
This work&
$256^3$&
Mat.&
$1.0$&
$0.25$&
$1.0$&
10&
128&
$12.0$&
100&
$0.0681\pm  0.0032$\tabularnewline
\hline
3D&
This work&
$512^3$&
Rad.&
$1.0$&
$0.25$&
$1.0$&
10&
256&
25.0&
10&
$0.0343\pm0.0023$\tabularnewline
\hline
3D&
This work&
$512^3$&
Mat.&
$1.0$&
$0.25$&
$1.0$&
10&
256&
25.0&
10&
$0.0468\pm0.0019$\tabularnewline
\hline\hline
4D&
 This work&
 $128^{4}$&
 Rad.&
 $0.1$&
 $0.05$&
 $1.0$&
 10&
 51&
 $5.0$&
 5&
 $0.0727\pm0.0013$\tabularnewline
\hline
4D&
 This work&
 $128^{4}$&
 Mat.&
 $0.1$&
 $0.05$&
 $1.0$&
 10&
 51&
 $5.0$&
 5&
 $0.0707\pm0.0005$\tabularnewline
\hline
4D&
This work&
$128^4$&
Rad.&
$1.0$&
$0.25$&
$1.0$&
10&
64&
$6.0$&
8&
$0.0743\pm0.0025$\tabularnewline
\hline
4D&
This work&
$128^4$&
Mat.&
$1.0$&
$0.25$&
$1.0$&
10&
64&
$6.0$&
8&
$0.0702\pm0.0044$\tabularnewline
\hline
\end{tabular}
\caption{\label{tab1}The measured scaling exponent $\lambda$ for $D=2$, $D=3$ and $D=4$ numerical simulations of domain wall networks. For comparison, we include both our own results and those of previous authors. The fourth column describes the cosmological epoch in which each set of simulations was performed. The ninth and tenth columns show respectively the \textit{total} conformal time dynamic range of the simulation (or the range until the horizon reaches half the box size, for the case of simulations that extend beyond this) and the 
conformal time dynamic range of the part of the simulation that was actually used in order to fit for the scaling exponent. In some cases explicit numbers are not provided by authors; whenever possible we have estimated them.}
\end{table*}

Our results are summarized in Table \ref{tab1}, where for comparison purposes we also show a number of previously obtained results. In each case we ran a number of different simulations, calculated an exponent for each simulation, and then computed the standard deviation for $\lambda$ from the ensemble of boxes. Note that other authors calculate errors in slightly different ways. For example, \cite{Garagounis} first carries out an average of all runs and then does the fit on the `averaged run'.

We quote both the full dynamic range (in conformal time) of each simulation and also the dynamic range corresponding to the interval that was actually used to fit for the scaling exponent. In some cases explicit details are not provided by previous authors and we have estimated them, either based on indirect information or by inspection, \textit{e.g.} of figures. For the case of \cite{Garagounis} the full dynamic range is effectively over-estimated, since early in their simulations dissipation was introduced in order to speed up the formation of the domain wall network. It is noticeable that larger, longer simulations do not necessarily correspond to a bigger dynamic range for the fit. Admittedly, this reflects the fact that different authors have different criteria to decide when they trust their simulations to be near the scaling regime, and hence which range to use in the fits. By comparison with previous work, our method seems to be fairly conservative, and this is one of the reasons why we can get quite small error bars. We have verified through simple tests (\textit{e.g.}, changing the ten percent cutoff to three percent) that within error bars our results do not critically depend on it.

Our results confirm previous work in that the approach to scaling is indeed very slow. We again find hints for deviations from the scale-invariant ($\lambda=0$) behaviour, which are consistent with most previous results (in particular with the original runs of Press, Ryden and Spergel \cite{Press}). This is to be contrasted this with the evolution of cosmic string networks, where field theory simulations with comparable or even smaller dynamic range find a relatively faster relaxation to scaling \cite{Moore,Testing}. On the other hand, notice that in the case of Garagounis and Hindmarsh \cite{Garagounis} no deviations from the scale-invariant behaviour are found in their standard simulations, but a one-off very large 3D simulation does find them. We do agree with \cite{Garagounis} that there is not any strong evidence for a logarithmic correction to the scaling law (which has been suggested by \cite{Press}), though it must be kept in mind that even with our larger dynamic ranges it would not be trivial to identify such a correction and, more to the point, to distinguish it from a power-law deviation to scale invariance.

Moreover, we caution that comparisons between \cite{Garagounis} and the other simulations are not straightforward for various reasons. Apart from their choice of physical rather than conformal time to present the results, they also use a different algorithm for calculating the domain wall areas and, more importantly, dissipation is put into the early part of the simulations. While this has the  obvious benefit of speeding up the formation of the wall network, and hence to some extent the approach to whatever is the late-time attractor solution for the network, it is not clear that it will not change the scaling properties themselves. At a qualitative level, we would expect that his could be the difference between their results and the others. Also, for a generic discussion of the possible pitfalls of lattice-based simulations see \cite{Scherrer}.

\begin{figure}
\includegraphics[width=3.5in]{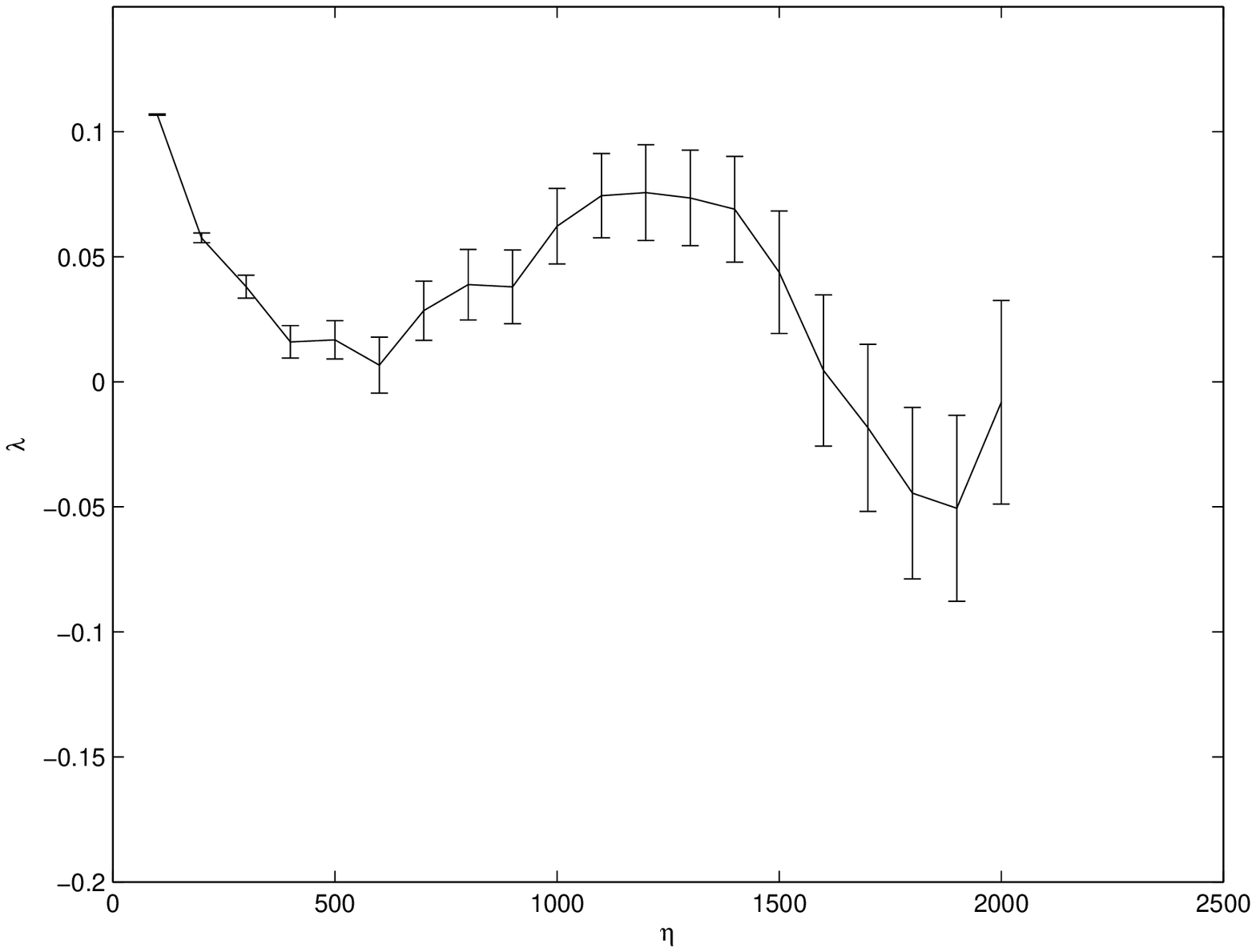}
\includegraphics[width=3.5in]{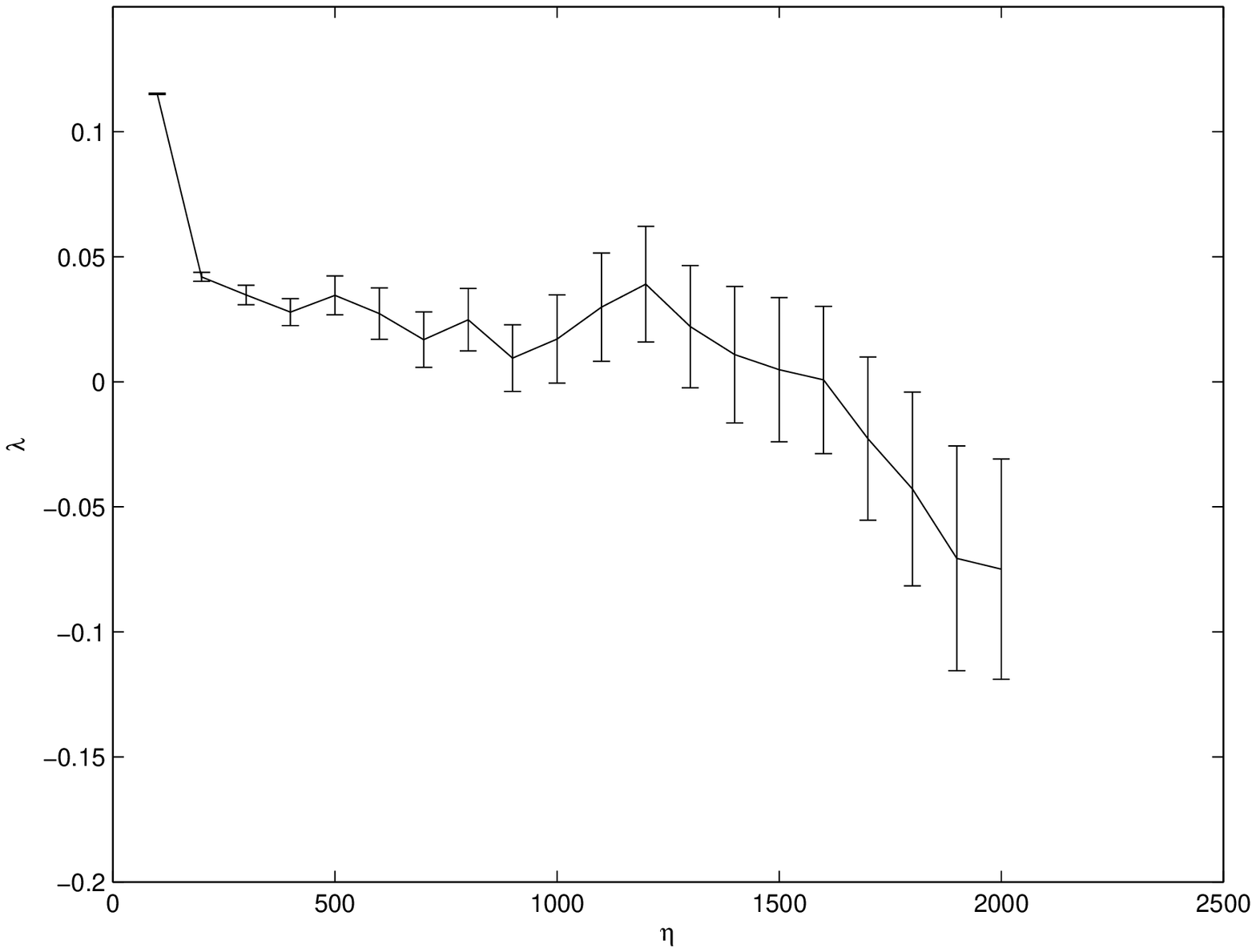}
\caption{\label{fig5}The evolution of the scaling exponent $\lambda$ with conformal time for 100 $4096^{2}$ simulations in the radiation (top) and matter (bottom) epochs. The dynamical range of each simulation was divided into 20 equally spaced bins, and an exponent measured for each bin and then averaged over the 100 runs. The error bar is the standard deviation in the ensemble.}
\end{figure}

Since our simulations have a large dynamical range, one can make the point that having a single exponent to characterize the entire range might not be a very good approximation, for presumably if the network is indeed approaching a linear scaling regime then the deviations should be decreasing as the simulations evolve. This is indeed a valid objection, and a simple way of studying what happens is as follows. We have taken our sets of 100 simulations of $4096^{2}$ and $256^{3}$ boxes and divided the dynamic range of each into a number of equally spaced bins in conformal time (20 bins for the former, 12 for the latter). We have then calculated a scaling exponent for each bin and run, and finally averaged over the ensemble of 100 runs for each of the bins. 

\begin{figure}
\includegraphics[width=3.5in]{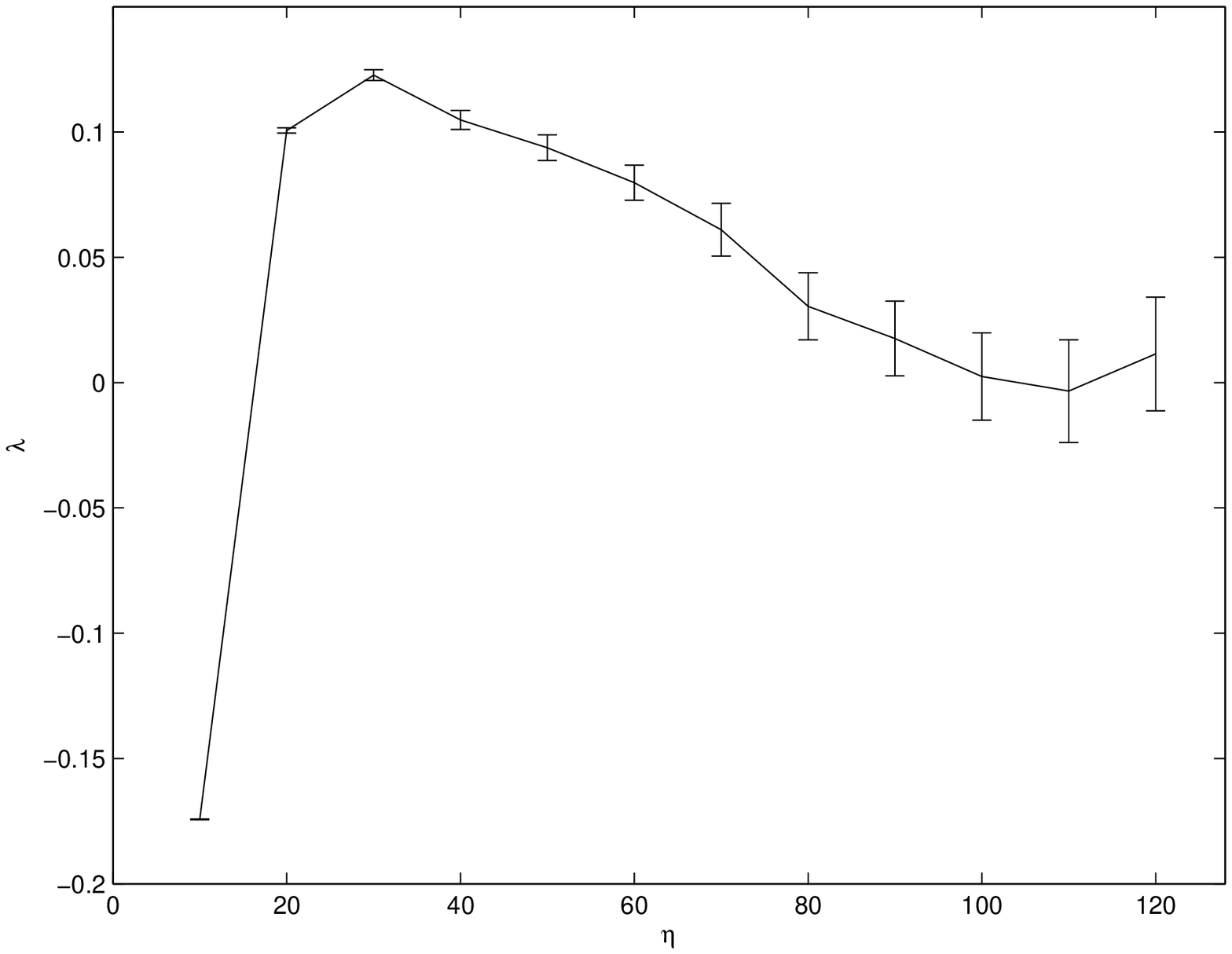}
\includegraphics[width=3.5in]{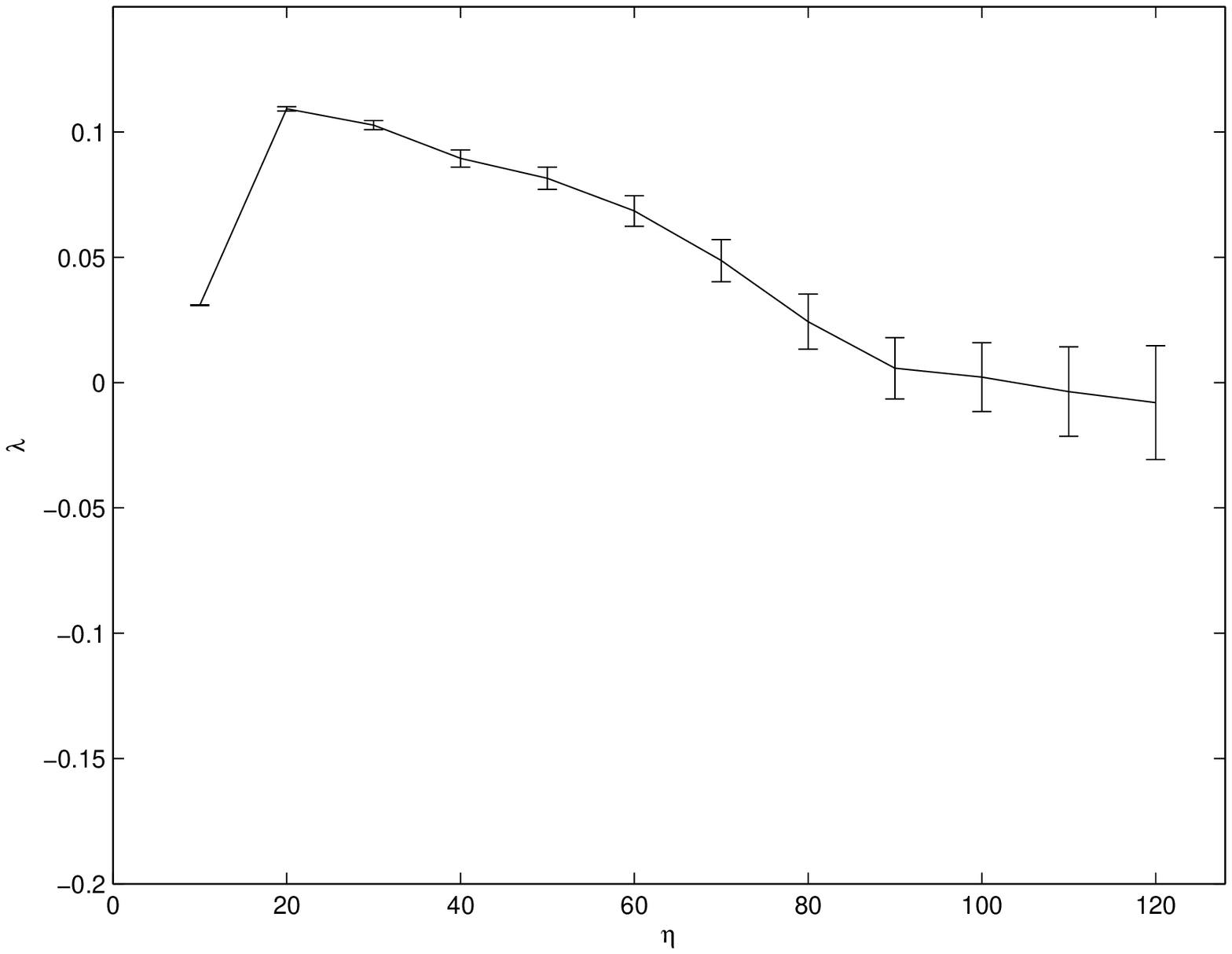}
\caption{\label{fig6}The evolution of the scaling exponent $\lambda$ with conformal time for 100 $256^{3}$ simulations in the radiation (top) and matter (bottom) epochs. The dynamical range of each simulation was divided into 12 equally spaced bins, and an exponent measured for each bin and then averaged over the 100 runs. The error bar is the standard deviation in the ensemble.}
\end{figure}

The results are shown in Figs. \ref{fig5} and \ref{fig6}. A noticeable feature is that the error bars tend to increase as the dynamic range increases. This is because as the Hubble volume becomes a larger fraction of the box size there are effectively less independent regions to sample from. The plots of the 3D simulations do show some evidence of evolution of the scaling exponent, which (if one neglects the initial bin) starts out around $\lambda_i=0.1$ and then slowly but steadily decreases. The situation is not so simple in the 2D case. Here in fact the exponent starts out at approximately the same value---a consequence of the fact that the same algorithm is being used to generate the initial conditions. However, in the longer dynamic range available the exponent oscillates and eventually `overshoots' the linear scaling regime, reaching negative exponents towards the end of the runs. Whether this is a numerical artifact or has some physical significance is presently unclear.

Notice that the fact that the exponent is negative (that is, $\xi$ is growing proportionally faster than time) does not necessarily mean there is any violation of causality. By this time there can be a sufficiently large number of walls inside a Hubble volume for such a regime to ensue, albeit transiently. A similar situation happens with the Kibble regime for the evolution of cosmic string networks \cite{ms1,ms2}. There, the network must be friction-dominated for such a regime to occur. In the present context this is not the case, though it can't be excluded that with the increasing dynamic range a background of small-scale features accumulates which could slow the walls down. Note that measuring velocities in field theory simulations is notoriously difficult---see \cite{Moore} for a discussion of the issue in the context of cosmic string simulations.

Finally, it must be emphasized that if the network does indeed not have a scale-invariant solution, or if it does have one but it takes much longer to reach it than the dynamic range available to existing simulations, then the initial conditions with which one start the simulations will of course play an important role in determining the subsequent evolution of the networks, and may therefore affect the scaling exponents obtained. This also needs to be taken into account when comparing the results obtained by different authors. Understanding the specific differences between the different available numerical pipelines, and possibly even carrying out independent tests with other algorithms are clearly key issues, and deserve further investigation. 

\section{\label{conc}Discussion and Conclusions}

We have presented the results of high-resolution, long dynamic range field theory simulations of domain wall networks. In addition to the usual 2D and 3D numerical simulations, we have also performed some simulations of domain wall evolution in four spatial dimensions. These can be relevant, in a phenomenological way, for models with additional space-time dimensions, such as the so-called brane world scenarios. 

Our results are further evidence of the fact that these networks have a very slow relaxation towards the expected asymptotic linear scaling solution. Even with the longest dynamic ranges we can presently run, we can not find unambiguous evidence for the onset of a linear scaling solution. The large number and size of our simulations has allowed us to measure scaling exponents with considerably smaller error bars than in previous works, thereby strengthening the evidence for small violations of scale invariance in the ranges we can probe. However, one is always limited by the finite dynamical range of our simulations and it is always possible that future simulations with a much larger dynamical range will produce results consistent with scaling. It is also worth emphasizing that the measured deviations tend to decrease when we increase the box size. Box size effects, as well as possible dependencies on the way initial conditions are generated, clearly deserve further investigation.

It is interesting to try to understand what could be the physical mechanism behind the violations to the scale invariance. Notice that typically they are such that the field configurations are equilibrating somewhat more slowly than would be allowed by causality (with the exception discussed above). As was already pointed out, measuring velocities is a very difficult task in this type of simulation, but we do find that our networks are relativistic. A possible explanation for the slower evolution is that it is due to the existence of long-range forces between the walls. However, if that was the case we would expect that the exponents $\lambda$, measuring the deviations from scale invariance, would decrease as the number of spatial dimensions increases: the more dimensions one has the easier it is for the walls to unwind, and the smaller (in relative terms) will be the effect of these forces. In fact, the opposite is the case in our simulations, although this is compounded by the fact that the dynamic ranges we can run are obviously smaller for larger dimensions, so one is not really comparing like with like.

Finally, let us point out that the existence of violations to scale invariance can have some cosmological consequences. Consider a network formed as early as observationally allowed, that is (roughly speaking, for the simplest standard scenarios) when the temperature of the Universe (photon temperature) was $T \sim 0.1 {\rm MeV}$. At formation there would be on average one domain per horizon volume, and that would also be the case today if the network quickly reached the linear scaling regime. However, if we assume that from then on $\xi_{c}=\eta^{1-\lambda}$ with $\lambda\neq0$ then the number of walls per horizon volume today could be as high as about ten for allowed values of the exponent. In other worlds, each Hubble volume could be divided into a significant number of domains, each conceivably having different properties. An example of such a model is described in \cite{Inhomog}. Although these would be important modifications to the standard domain wall scenario we do not expect the limits on the wall mass per unit area (or equivalently to the symmetry breaking scale) to be severely modified.

\section{acknowledgments}
We thank Ruth Durrer and Paul Shellard for useful discussions and 
suggestions. J.O. is grateful for the hospitality of DAMTP (Cambridge), where some of the present work was carried out.
This work was done in the context of the ESF COSLAB network, and was performed
on COSMOS, the Altix3700 owned by the UK Computational Cosmology Consortium, supported by SGI, Intel, HEFCE and PPARC. 


\bibliography{scaling}

\end{document}